\documentclass[12pt]{article}
\usepackage{amsmath,amssymb,graphicx,subfigure}
\makeatletter \@addtoreset{equation}{section}

\usepackage{cite}
\usepackage{bm}
\usepackage{dcolumn}
\makeatother
\def\one{{\hbox{ 1\kern-.8mm l}}}
\newcommand{\Dslash}{\not{\hbox{\kern-4pt $D$}}}
\newcommand{\pdslash}{\not{\hbox{\kern-2pt $\partial$}}}
\newcommand{\be}{\begin{equation}}
\newcommand{\bea}{\begin{eqnarray}}
\newcommand{\eea}{\end{eqnarray}}
\newcommand{\ba}{\begin{array}}
\newcommand{\ea}{\end{array}}
\newcommand{\ee}{\end{equation}}
\newcommand{\nn}{\nonumber}

\expandafter\ifx\csname mathbbm\endcsname\relax

\newcommand{\ds}{\!\!\!\!\! \hspace{1mm} /}
\else

\fi \textheight 22cm \textwidth 15cm \topmargin 1mm
\begin{document}

\begin{titlepage}
\hfill
\vbox{
    \halign{#\hfil         \cr
           IPM/P-2012/003 \cr
                      } 
      }  
\vspace*{20mm}
\begin{center}
{\Large {\bf Fermions on Lifshitz Background}\\
}

\vspace*{15mm}
\vspace*{1mm}
{Mohsen Alishahiha, M. Reza Mohammadi Mozaffar and  Ali Mollabashi }

 \vspace*{1cm}

{\it  School of physics, Institute for Research in Fundamental Sciences (IPM)\\
P.O. Box 19395-5531, Tehran, Iran \\ }

\vspace*{.4cm}

{E-mails: {\tt alishah, m$_{-}$mohammadi, mollabashi@ipm.ir}}%

\vspace*{2cm}
\end{center}

\begin{abstract}

We study  a non-relativistic fermionic retarded Green's function by making use of a fermion on the Lifshitz geometry with critical exponent $z=2$.
With a natural boundary condition, respecting the symmetries of the model, the resultant retarded Green's function exhibits a number of interesting features including a flat band. We also study the finite temperature and finite chemical potential cases where the geometry is replaced by  Lifshitz black hole solutions.
\end{abstract}

\end{titlepage}


\section{Introduction}

At critical points, physics is usually described by a scale invariant model. Typically
the scale invariance arises in the relativistic conformal group where we have
\be
t\rightarrow \lambda t,\;\;\;\;\;\;x_i\rightarrow \lambda x_i.
\ee
Here $t$ is time and $x_i$'s are spatial directions of the space time.

We note, however, that in many physical systems the critical points are governed by
dynamical scalings in which space and time scale differently. In fact  spatially
isotropic scale invariance is characterized by the dynamical exponent, $z$,
 as follows \cite{Hertz:1976zz}
\be\label{Lif}
t\rightarrow \lambda^zt,\;\;\;\;\;\;x_i\rightarrow \lambda x_i.
\ee
The corresponding critical points are known as Lifshitz fixed points.

In light of AdS/CFT correspondence \cite{Maldacena:1997re} it is natural to seek for
gravity duals of Lifshitz fixed points. Indeed  gravity descriptions of
Lifshitz fixed points have been  considered in\cite{Kachru:2008yh}(see also \cite{Koroteev:2007yp}
for an earlier work on a geometry with the Lifshitz scaling,) where
a metric  invariant under the scaling \eqref{Lif} was introduced.
The corresponding metric is\footnote{As it has been mentioned in \cite{Kachru:2008yh},
although the metric is nonsingular,
it is not geodesically complete and in particular an infalling object into $r=\infty$  feels
a large tidal force.}
\be\label{BacLif}
ds^2=L^2\left(- \frac{dt^2}{r^{2z}}+\frac{d\vec{x}^2}{r^2}+\frac{dr^2}{r^2}\right),
\ee
where $L$ is  the radius of curvature\footnote{In what follows we set $L=1$.}.  The action of the scale transformation \eqref{Lif} on the metric is given by
\be
t\rightarrow \lambda^zt,\;\;\;\;\;\;x_i\rightarrow \lambda x_i,\;\;\;\;\;\;
r\rightarrow \lambda^{-1}r.
\ee

As a physical application, the Lifshitz geometry has been used to provide a  possible
holographic description  for strange metals\cite{Hartnoll:2009ns}\footnote{
See also \cite{Fadafan:2009an} for drag force computations in Lifshitz
geometries.}. In this setup the Lifshitz
background is probed by D-branes with non-zero gauge fields in their  world
volume. By choosing the dynamical critical exponent, $z$, the authors
of \cite{Hartnoll:2009ns}
have been able to match the non-Fermi liquid scalings, such as linear resistivity, observed in
strange metal regimes. Having found the non-Fermi liquid scalings, it is natural to study the fermionic properties of the system to explore, for example, a possibility of having a Fermi surface in the model. To do so, one needs to consider a fermion
on the Lifshitz geometry to find the retarded Green's function of the
corresponding dual fermionic operator via AdS/CFT correspondence.

Indeed utilizing fermions on asymptotically AdS geometries, it was shown that the AdS/CFT correspondence can holographically describe
Fermi surfaces \cite{{Lee:2008xf} ,{Liu:2009dm},{Cubrovic:2009ye},{Faulkner:2009wj},{Faulkner:2010zz} ,{Hartnoll:2011dm},{Cubrovic:2011xm}}.
Actually to see a Fermi surface, one should look
for a sharp behavior in the fermionic retarded Green's function at
finite momentum and small frequencies (for a review see e.g.\cite{Faulkner:2011tm}).
Moreover  the spectrum of quasi-particle
excitations near the Fermi surface is governed by an emergent CFT corresponding to
the AdS$_2$ near horizon geometry of the  black hole\cite{Faulkner:2009wj}.

The aim of this article is  to study fermions on the Lifshitz geometry which in turns
can be used to study the  fermionic retarded Green's function of the corresponding
non-relativistic dual theory\footnote{Fermions on Schrodinger spacetime has also been
studied in\cite{{Akhavan:2009ns},{Leigh:2009ck}}.}. Fermions on the Lifshitz geometry or on a geometry with emergent Lifshitz geometry in its
near horizon limit have been studied in \cite{korovins} and \cite{Faulkner:2010tq}
respectively. In these papers, with a Lorentz symmetric 
 boundary term, the real time Green's function of the
fermion has been obtained. It was shown that the resultant Green's function has real self energy.
We will come back to this point in section two.

In this paper in order to find the retarded Green's function we consider the
Lorentz symmetry breaking boundary condition introduced in \cite{Tong}\footnote{
Throughout this paper we refer to this boundary condition as {\it non-standrad}
boundary condition, while we refer to that introduced in \cite{{Henningson:1998cd},{Henneaux:1998ch}} as {\it standard} boundary condition.}. Since
the corresponding boundary condition preserves rotational and scale invariances, but
breaks the boost, it is more natural to impose such a boundary condition on
the geometries with  Lifshitz isometry. Of course it also breaks the parity which
is preserved by the Lifshitz symmetry\footnote{The Lorentz symmetry breaking
boundary condition has also been imposed  for  fermions
with dipole coupling in \cite{Li:2011nz} and, on the
 charged dilatonic black hole in \cite{Li:2011sh}.}.

 Note that in this paper we consider the fermion as a probe. It would be
interesting to extend this work to the case where the back reaction of the fermions is
taken into account.

The paper is organized as follows. In the next section we study fermions on the
Lifshitz geometry where we find a solution for the equation of motion with a
proper  boundary condition. Then using the solution we calculate the corresponding
retarded Green's function where we see that the model exhibits a flat band. In section
three we extend our study to the finite temperature case where we show that
although the system has excited zero energy fermionic modes at low momenta, at high
momenta still it has a flat band.
In section four we consider  charged fermions probing a charged Lifshitz black hole where we show that while
with the standard boundary condition the system exhibits a Fermi surface, in the
non-standard case it still has flat band. The last section is devoted to discussions.


\section{Zero temperature}

The aim of this  section is to study fermions on the Lifshitz background which will be used
to find the retarded Green's function for the corresponding fermionic dual operator
in the dual non-relativistic field theory. Before going into computations, it is worth to
note that the Lifshitz geometry is not a solution of the pure Einstein gravity with or without
cosmological constant.

In general to get the Lifshitz geometry one needs to couple the Einstein gravity
to other fields. In particular the Lifshitz geometry may be obtained from gravity coupled to massive gauge fields.  In the minimal case where we have only one  massive gauge field,
the corresponding action is given as follows
\be
I=\frac{1}{16\pi G_{d+1}}\int d^{d+1}x\;\sqrt{g}\;\left(R+\Lambda-\frac{1}{4}F^2
-\frac{1}{4}m^2A^2\right),
\ee
where $F_{\mu\nu}=\partial_\mu A_\nu-\partial_\nu A_\mu$.
 It is easy to
see that, with a suitable choice of the parameters $m$ and $\Lambda$,  the model admits
the Lifshitz solution \eqref{BacLif}  with a non-zero gauge field given by \cite{Taylor:2008tg}
\be
A_t=\sqrt{\frac{2(z-1)}{z}}\frac{1}{r^{z}}.
\ee
For this solution the parameters $m$ and $\Lambda$ are
$m^2=4z,\;\Lambda=z^2+(d-2)z+(d-1)^2$.

Alternatively
Lifshitz metric may also be obtained as
a solution of the pure gravity modified by curvature squared terms\cite{AyonBeato:2010tm}.
 As the simplest case consider a $d+1$ dimensional gravitational action as follows
 \be I = \frac{1}{2\kappa^2}\,
  \int d^{d+1}x\sqrt{-g}\;(R -\Lambda+ \beta  R^2 ) \,.\label{action}
\ee
Using the equations of motion derived from the above action
one can show that the Lifshitz geometry \eqref{BacLif} is a solution of the equations of motion  for a suitable choice of the cosmological constant
$\Lambda$ and the coupling constant
$\beta$  that are given by
\be
\Lambda=-\frac{2z^2+(d-1)(2z+d)}{2},\;\;\;\;\;\;\;\beta=-\frac{1}{4\Lambda}.
\ee

Although we could have Lifshitz metric in arbitrary dimensions, in what follows we will
consider the four dimensional Lifshitz geometry which could provide a holographic
description for a three dimensional non-relativistic field theory.


\subsection{Fermions on Lifshitz geometry}

Let us consider a four dimensional Dirac fermion on the Lifshitz background whose
action is
\be S_{\rm bulk} =\int d^4x \sqrt{-g}\,i\bar{\Psi} \,\left[\frac{1}{2}\left(\Gamma^a\!\stackrel{\rightarrow}D_a-
\stackrel{\leftarrow}D_a\!\Gamma^a\right)-m\right]\Psi\label{Action}\ee
where
${D\ds}=(e_\mu)^a\Gamma^\mu[\partial_a+\frac{1}{4}(\omega_{\rho\sigma})_a\Gamma^{\rho\sigma}]
$, with
$\Gamma^{\mu\nu}=\frac{1}{2}[\Gamma^\mu,\Gamma^\nu]$. In our notation the
space-time indices are denoted by $a,b\cdots$, though the tangent space indices
are labeled by $\mu,\nu\cdots$.

Since the Lifshitz metric may be obtained from a gravity coupled to a massive
gauge field, in general, the solution may also support a non-zero gauge field. Therefore
one could consider a fermion that is charged under the background gauge field. Nevertheless
in what follows we will consider a neutral fermion.  We will come  back to  charged fermions, latter, in the discussion section.

To write the equation of motion one should use the variational principle which
typically comes with a proper boundary condition. It is important to note that the boundary term is not unique and indeed there are several
ways to make the variational principle well defined using different boundary terms \cite{Tong}. For the moment, we assume that there is
a suitable boundary condition such that the variational principle will be well defined.
With this assumption the equation  of motion is
\be
\left((e_\mu)^a\Gamma^\mu[\partial_a+\frac{1}{4}(\omega_{\rho\sigma})_a\Gamma^{\rho\sigma}]-m\right)\Psi=0,
\ee
where  the nonzero components of vierbeins and spin connections
for the  Lifshitz metric \eqref{BacLif} are
\begin{eqnarray*}
(e_t)^a=r^z\delta^{ta},\hspace{5mm}
(e_i)^a=r\delta^{ia},\hspace{5mm}
(e_r)^a=r\delta^{ra},
\end{eqnarray*}
and
\begin{eqnarray*}
(\omega_{tr})_a=-(\omega_{rt})_a=\frac{z}{r^z}\delta_{ta},\hspace{5mm}
(\omega_{ir})_a=-(\omega_{ri})_a=-\frac{1}{r}\delta_{ia}.
\end{eqnarray*}
Using these expressions the equation of motion reduces to
\begin{eqnarray}
\left[\Gamma^tr^z\partial_t-\left(\frac{z}{2}+1\right)\Gamma^r+r\Gamma^i\partial_i+r\Gamma^r\partial_r-m\right]\Psi=0.
\end{eqnarray}
To proceed it is useful to work in the momentum space where we may set  $\Psi=e^{i\omega t+ik.x}\psi(r)$. In this notation the  equation of motion reads
\be\label{firstOrderResult}
ir(k.\Gamma)\psi=\left[-i\omega r^z\Gamma^t+\left(\frac{z}{2}+1\right)\Gamma^r-r\Gamma^r\partial_r+m\right]\psi.
\ee
It is also useful to act by $(D\ds+m)$ on the first order equation of motion to find a second order
differential equation which typically is easer to solve. Doing so, and using
equation \eqref{firstOrderResult}, one arrives at
\bea
({D\ds}{D\ds}-m^2)\psi
&=&\bigg[r^2\partial^2_r-(z+2)r\partial_r+\omega^2 r^{2z}+\left(\frac{z}{2}+1\right)\left(\frac{z}{2}+2\right)-r^2\vec{k}^2\cr &&\;\;\;
+i(z-1)\omega r^z\Gamma^r\Gamma^t+m\Gamma^r-m^2\bigg]\psi=0.
\eea
In general this equation may not  have  analytic solutions. We note, however, that for a particular case of $m=0$ and $z=2$ the equation has, indeed, an analytic solution. This is the case we
will consider in this paper. In this case defining $\psi_\pm=\frac{1}{2}(1\pm \Gamma^r\Gamma^t)\psi$,
one gets
\begin{eqnarray}\label{eq1}
\bigg[r^2\partial^2_r-4r\partial_r+\omega^2 r^{4}-r^2(\vec{k}^2\mp i\omega)
+6 \bigg]\psi_\pm&=&0.
\end{eqnarray}
To solve the above equation we make the following change of variable
\be
\psi_\pm(r)=r^{3/2} e^{\frac{i\omega}{2}r^2}f_{\pm}(i\omega r^2)
\ee
by which equation \eqref{eq1} reduces to a well known differential equation  for
$f_\pm(\xi)$
\be
\frac{d^2f_\pm(\xi)}{d\xi^2}+\frac{df_\pm(\xi)}{d\xi}+\left(\frac{\lambda_\pm}{\xi}+
\frac{\frac{1}{4}-\mu_\pm^2}{\xi^2}\right)f_\pm(\xi)=0,
\ee
where
\be
\lambda_{\pm}=-\frac{k^2}{4i\omega}\pm\frac{1}{4},\;\;\;\;\;\;\;\;\;
\mu_\pm=\frac{1}{4}.
\ee
We recognize the above equation as the hypergeometric differential equation whose solution
is
\be
f_{\pm}(\xi)=c^\pm_1 \xi^{\frac{1}{2}-\mu_\pm}e^{-\xi}
F(\alpha_\pm,-2\mu_\pm+1,\xi)+c^\pm_2 \xi^{\frac{1}{2}+\mu_\pm}e^{-\xi}
F(\beta_\pm,2\mu_\pm+1,\xi),
\ee
where $F(a,b,\xi)$ is the confluent hypergeometric function, $c^{\pm}_{1,2}$ are two constant spinors and
\be
\alpha_\pm=\frac{1}{2}-\mu_\pm-\lambda_\pm,\;\;\;\;\;\;\;\;\;\;\;\;\;
\beta_\pm=\frac{1}{2}+\mu_\pm-\lambda_\pm.
\ee
Therefore altogether we find
\be
\psi_\pm(r)=e^{-\frac{i\omega}{2}r^2}r^2\bigg[D^{\pm}_1
F\left(\alpha_\pm,\frac{1}{2},i\omega r^2\right)+D^{\pm}_2r
F\left(\beta_\pm,\frac{3}{2},i\omega r^2\right)\bigg],
\ee
with $D^\pm_1=(i\omega)^{\frac{1}{4}}c_1^\pm$,
$D^\pm_2=(i\omega)^{\frac{3}{4}}c_2^\pm$.

It is important to note that so far we have solved the second order differential equation and thus
the constant spinors $c_{1,2}^{\pm}$ are not independent and, indeed, restricting
the above solution to be a solution of the first order equation of motion \eqref{firstOrderResult} leads to certain relations among them. More precisely one finds
\begin{eqnarray}\label{res2}
c_2^+=\frac{-i}{\sqrt{i\omega}}\Gamma^r (k.\Gamma) c_1^-,
\hspace{1.5cm}
c_2^-=\frac{-i}{\sqrt{i\omega}}\Gamma^r (k.\Gamma) c_1^+.
\end{eqnarray}

We note, also, that the solution has not been uniquely fixed yet. In fact in the context of AdS/CFT correspondence one usually imposes a boundary condition
at IR. In the Euclidean case the proper boundary condition is to assume that
the wave function is finite at IR.  When we are dealing
with the real-time AdS/CFT correspondence, the proper boundary condition is to
impose an ingoing boundary condition on the  wave function  at the horizon \cite{Son:2002sd}.
 In our case using the asymptotic behavior of the hypergeometric function,
\be
F(a,b,\xi)\sim \frac{\Gamma(b)}{\Gamma(b-a)}(-\xi)^{-a}+\frac{\Gamma(b)}{\Gamma(a)}
e^{\xi}\xi^{a-b},\;\;\;\;\;\;\;\;\;{\rm for\; large}\;|\xi|,
\ee
the wave function is ingoing at ``Lifshitz horizon'' at $r=\infty$,  if the parameters $c_1^+$ and
$c_2^+$ satisfy the following relation
\be\label{eq-in}
c^+_2=-2
\frac{\Gamma(\alpha_++\frac{1}{2})}{\Gamma(\alpha_+)}\;c^+_1,
\ee
by which the ingoing wave function near the Lifshitz horizon behaves as follows\footnote{Since in what follows we are
interested in the low energy limit of the retarded Green's function, the momentum will be space like, {\it i.e.}
$k^2\leq \omega^2$.}
\be
\psi\sim r^2 e^{i\omega(t-\frac{r^2}{2}+\frac{k^2}{2\omega^2}{\rm ln}r)}.
\ee 

One may wonder what ``Lifshitz horizon'' means! Actually the situation
could be compared  with that of fermions on the pure AdS case studied in, e.g.    \cite{Iqbal:2009fd} where the ingoing boundary condition has been imposed at
the AdS horizon at $r=\infty$. In order to understand it  better one may think of this
boundary condition as a limiting procedure starting from a black hole solution and then
approaching the zero temperature limit, as we will do in the next section.

Alternatively to obtain the relation \eqref{eq-in} and then the
corresponding  retarded Green's function\footnote{The
prescription for calculating retarded Green's function in the context of
AdS/CFT correspondence has been first considered in \cite{Son:2002sd}
 and further studied in the literature in, e.g. \cite{{Herzog:2002pc},{Marolf:2004fy},{Gubser:2008sz},{Skenderis:2008dg},{vanRees:2009rw},{Iqbal:2008by}}.}  one
may use the prescription explored in  \cite{Iqbal:2009fd} where the authors
presented a derivation of the real-time AdS/CFT prescription as an analytic continuation
of the corresponding problem in the Euclidean signature. Indeed in our case,
we have checked that, using this
prescription  we
will arrive at the same results as those in this and the next subsections\footnote{Euclidean
Green's function for fermions on the Lifshitz geometry has recently been studied in
\cite{korovins}.}.

\subsection{Retarded Green's function}

In this subsection we  compute the retarded Green's function
of a fermionic operator in the dual non-relativistic three dimensional
field theory by making use of the
solution we obtained in the previous section. One should note that, in the context of the
AdS/CFT correspondence,
in order to find the corresponding retarded Green's function it is crucial to appropriately identify
the source and response of the dual operator.

 On the other hand the identification of the  source and response
depends on the boundary conditions which one imposes to
get a well defined variational principle. Thus it is important to
study the possible boundary terms one may add to the action to make the
variational principle well defined. Therefore in what follows we will first find a proper boundary action for our model. To
do so it is useful to explicitly fix our notation.

Since we have been working in a basis in which $\Gamma^r\Gamma^t$ is diagonal,
we  use the following
representation for four dimensional gamma matrices
\begin{eqnarray}\label{basis}
\Gamma^r=
\begin{pmatrix}
-\sigma^2&0\\0&\sigma^2
\end{pmatrix},
\Gamma^t=
\begin{pmatrix}
i\sigma^1&0\\0&i\sigma^1
\end{pmatrix},
\Gamma^1=
\begin{pmatrix}
-\sigma^3&0\\0&-\sigma^3
\end{pmatrix},
\Gamma^2=
\begin{pmatrix}
0&-i\sigma^2\\i\sigma^2&0
\end{pmatrix}.
\end{eqnarray}
In this notation one has
\begin{eqnarray}\label{psi+}
\Psi_+&=&\frac{1}{2}\left(1+\Gamma^r\Gamma^t\right)\Psi
\,\,\,=\,\,\,
\mathrm{diag}(0,1,1,0)
\begin{pmatrix}
\Psi_1\\\Psi_2\\\Psi_3\\\Psi_4
\end{pmatrix}
\,\,\,=\,\,\,
\begin{pmatrix}
0\\\Psi_2\\\Psi_3\\0
\end{pmatrix}
\\
\Psi_-&=&\frac{1}{2}\left(1-\Gamma^r\Gamma^t\right)\Psi
\,\,\,=\,\,\,
\mathrm{diag}(1,0,0,1)
\begin{pmatrix}
\Psi_1\\\Psi_2\\\Psi_3\\\Psi_4
\end{pmatrix}
\,\,\,=\,\,\,
\begin{pmatrix}
\Psi_1\\0\\0\\\Psi_4
\end{pmatrix}\label{psi-}
\end{eqnarray}
Therefore the boundary terms coming from the variation of  the bulk action,
\be
\delta S_{\rm bulk}=\frac{i}{2}\int d^3x \sqrt{-h} (\bar{\Psi}\Gamma^r\delta\Psi-\delta\bar{\Psi}
\Gamma^r\Psi),
\ee
reads
\bea
&&\delta S_{bulk}=\frac{i}{2}\int d^3x\sqrt{-h}\;
\bigg[\Psi^{\dagger}_1\delta\Psi_1-\Psi^{\dagger}_2\delta\Psi_2-\Psi^{\dagger}_3
\delta\Psi_3+\Psi^{\dagger}_4\delta\Psi_4\cr
&&\;\;\;\;\;\;\;\;\;\;\;\;\;\;\;\;\;\;\;\;\;\;\;\;\;\;\;\;\;\;\;\;\;\;\;\;\;
-\delta\Psi^{\dagger}_1\Psi_1+\delta\Psi^{\dagger}_2\Psi_2
+\delta\Psi^{\dagger}_3\Psi_3-\delta\Psi^{\dagger}_4\Psi_4
\bigg].
\eea

Since the Dirac equation is a first order differential equation we are not allowed to impose
the boundary condition on all components of the spinors. Thus the aim is to add
a proper boundary term such that half of the degrees of freedom do not appear on the
boundary. So  we will have to fix only half of the spinors.

We note, however, that the boundary terms may not be unique \cite{Tong}. Of course different boundary terms lead to different physics. In our case
since the dual theory is a non-relativistic field theory, one may relax the condition to have
Lorentz symmetric boundary terms.

Following the suggestion of \cite{Tong} it is natural to consider the following
boundary term\footnote{Note that this boundary term is different from that considered in \cite{korovins}.}
\be\label{boundary}
S_{bdy}=\frac{1}{2}\int d^3x\sqrt{-h}\bar{\Psi}\Gamma^1\Gamma^2\Psi
\ee
which in our notation reads
\be
S_{bdy}=\frac{i}{2}\int d^3x\sqrt{-h}(\Psi_1^\dagger\Psi_3+\Psi_2^\dagger\Psi_4-
\Psi_3^\dagger\Psi_1-\Psi_4^\dagger\Psi_2).
\ee
This boundary term is invariant under rotation and scaling but breaks the boost
symmetry. Of course in our model, being Lifshitz geometry, the boost symmetry
has already been broken by the geometry at first place.

\begin{figure}
\begin{center}
\includegraphics[height=4.5cm, width=5cm]{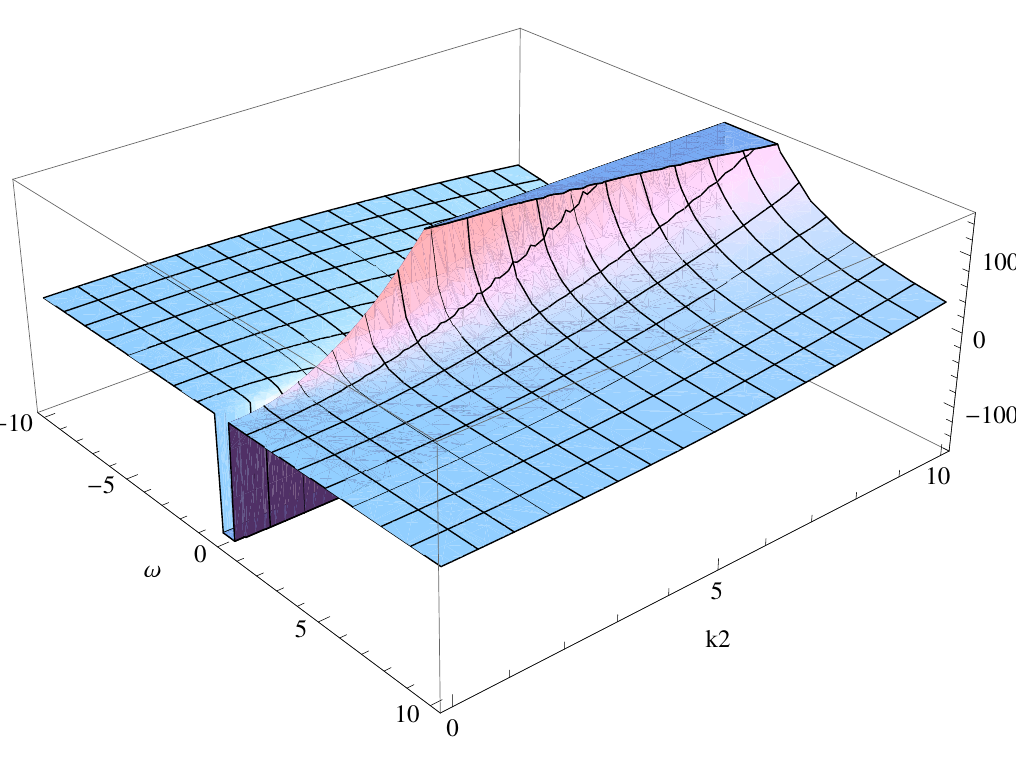}
\hspace{2cm}
\includegraphics[height=4.5cm, width=5cm]{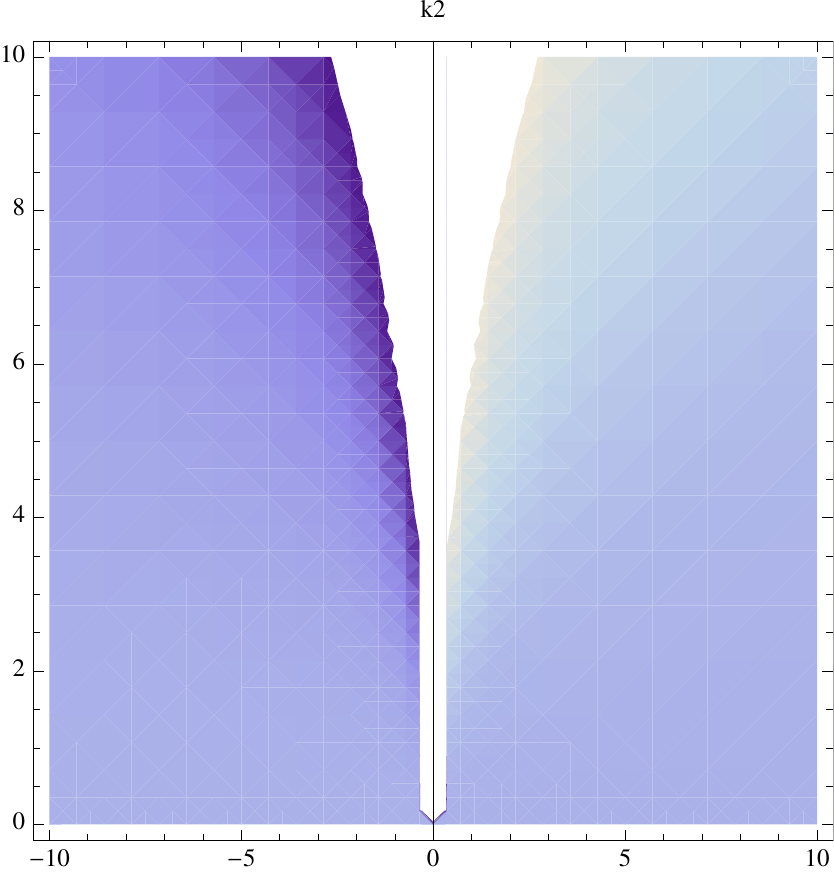}
\caption{Three dimensional and density plots of the spectral function
as a function of $\omega$ and $k_2$. It is positive for ${\rm sign}(\omega)>0$ and has a
pole
at $\omega=0$ for fixed $k_2$.} \label{fig1}
\end{center}
\end{figure}

Adding this boundary term to the bulk action and varying the total action we
arrive at
 \begin{eqnarray*}
\delta S_{bulk}+\delta S_{bdy}&=&\frac{i}{2}\int\sqrt{-h}\Big[\delta(\Psi^{\dagger}_1+\Psi^{\dagger}_3)(\Psi_3-\Psi_1)+\delta(\Psi^{\dagger}_2-\Psi^{\dagger}_4)(\Psi_2+\Psi_4)\\&&
\;\;\;\;\;\;\;\;\;\;\;\;\;\;\;\;\;+
(\Psi^{\dagger}_1-\Psi^{\dagger}_3)\delta(\Psi_1+\Psi_3)+(\Psi^{\dagger}_2
+\Psi^{\dagger}_4)\delta(\Psi_4-\Psi_2)\Big]
\\&=&
i\int\sqrt{-h}\left[-\delta \chi^{\dagger}_1\chi_2+\delta \zeta^{\dagger}_2\zeta_1+ \chi_2^{\dagger}\delta \chi_1-\zeta_1^{\dagger}\delta \zeta_2\right]
\end{eqnarray*}
where
\bea\label{difi}
(\chi_1,\chi_2)&=&\frac{1}{\sqrt{2}}(\Psi_1+\Psi_3,\Psi_1-\Psi_3)\cr
(\zeta_1,\zeta_2)&=&\frac{1}{\sqrt{2}}(\Psi_2+\Psi_4,\Psi_2-\Psi_4).
\eea
Therefore we get a well defined variational principle by setting a Dirichlet
boundary condition on $\chi_1$ and $\zeta_2$\footnote{If we had considered
$\delta S_{bulk}-\delta S_{bdy}$ the boundary condition should have been
imposed on $\chi_2$ and $\zeta_1$.}. As a result the source and response are given by
 $(\chi_1,\zeta_2)$ and $(\chi_2,\zeta_1)$, respectively. The
retarded Green's function is essentially a matrix which maps the source to the
response.
To compute the corresponding retarded Green's function it is illustrative to
explicitly  write the solution we have found in the previous section in
components
\be
\begin{pmatrix}
\psi_1\\ \\  \psi_2\\ \\  \psi_3\\ \\  \psi_4
\end{pmatrix}
=r^2e^{-\frac{i\omega}{2}r^2}
\begin{pmatrix}
D^-_{1\uparrow}F\left(\alpha_-;\frac{1}{2};i\omega r^2\right)+D^-_{2\uparrow}
 r F\left(\beta_-;\frac{3}{2};i\omega r^2\right)\\ \\
D^+_{1\downarrow}F\left(\alpha_+;\frac{1}{2};i\omega r^2\right)+D^+_{2\downarrow} r F\left(\beta_+;\frac{3}{2};i\omega r^2\right)\\ \\
D^+_{1\uparrow}F\left(\alpha_+;\frac{1}{2};i\omega r^2\right)+D^+_{2\uparrow}
 r F\left(\beta_+;\frac{3}{2};i\omega r^2\right)\\ \\
D^-_{1\downarrow}F\left(\alpha_-;\frac{1}{2};i\omega r^2\right)
+D^-_{2\downarrow} r F\left(\beta_-;\frac{3}{2};i\omega r^2\right)
\end{pmatrix}
\ee
On the other hand in our basis one has
\begin{eqnarray}
\Gamma^r(k.\Gamma)=
i\begin{pmatrix}
k_1\sigma^1&k_2\\k_2&-k_1\sigma^1
\end{pmatrix}
\end{eqnarray}
Therefore the equation  \eqref{res2} reads
\bea
\begin{pmatrix}
0\\
c_{2\downarrow}^+\\
c_{2\uparrow}^+\\
0
\end{pmatrix}
=\frac{1}{\sqrt{i\omega}}
\begin{pmatrix}
0\\
k_1c_{1\uparrow}^-+k_2c_{1\downarrow}^-\\
k_2c_{1\uparrow}^--k_1c_{1\downarrow}^-\\
0
\end{pmatrix}
=\frac{1}{\sqrt{i\omega}}
\begin{pmatrix}
0\\
Ac_{1\downarrow}^+\\
Ac_{1\uparrow}^+\\
0
\end{pmatrix}
\eea
where the last equality is the  ingoing condition \eqref{eq-in} with
\begin{eqnarray}\label{A}
A=-2\sqrt{i\omega}
\frac{\Gamma(\alpha_++\frac{1}{2})}{\Gamma(\alpha_+)}.
\end{eqnarray}
By making use of  this relation and utilizing the asymptotic behaviors of the
solution, the retarded
Green's function can be read
as follows
\begin{eqnarray}\label{G0}
G_R(k)=-\frac{1}{A^2-k^2}
\begin{pmatrix}
 A^2-2Ak_2+k^2&
 -2k_1A\\ &\\
-2 k_1A&A^2+2Ak_2+k^2
\end{pmatrix}.
\end{eqnarray}
As it is evident from the above expression for the retarded Green's function $G_{11}(\omega,k_2)=
G_{22}(\omega,-k_2)$ and $\det(G_R)=-1$. The spectral function is also
given by
\be
{\cal A}(k)=-\frac{1}{\pi}{\rm Im}({\rm Tr}(G_R))=\frac{2}{\pi}{\rm Im}
\bigg(\frac{A^2+k^2}{A^2-k^2}\bigg).
\ee
To study different features of  the retarded Green's function one may use the rotational
 symmetry  to set $k_1=0$\footnote{Although the model has rotational symmetry, the resultant
retarded Green's function \eqref{G0} seems asymmetric with respect to exchanging $k_1$ and
$k_2$. We note that it is the artificial of our asymmetric representation of the Gamma
matrices. We will back to this point latter.}. In this case the retarded Green's function
becomes diagonal. In fact taking into account that
 $k=k_2$ one gets
\begin{eqnarray}
G_R(k_2)=-
\begin{pmatrix}
\frac{A-k_2}{A+k_2}&
 0\\ &\\
0&\frac{A+k_2}{A-k_2}
\end{pmatrix}.
\end{eqnarray}
Moreover one finds
\be\label{A0}
{\cal A}(k_2)=\frac{2}{\pi}{\rm Im}
\left(\frac{\Gamma^2(\eta+\frac{1}{2})+\Gamma(\eta)\Gamma(\eta+1)}{
\Gamma^2(\eta+\frac{1}{2})-\Gamma(\eta)\Gamma(\eta+1)}\right),\;\;\;\;\;\;\;
{\rm with}\;\eta=\frac{k_2^2}{4i\omega}.
\ee
The behavior of the spectral function as a function of $k_2$ and $\omega$ is shown
in figure \ref{fig1}.  Since the spectral function is symmetric under $k\rightarrow -k$, it is
sufficient to draw the figure for positive $k$.  As we observe the spectral function is positive for ${\rm sign}(\omega)>0$ and diverges at $\omega\rightarrow 0$.
Actually by making use of the asymptotic behavior of the Gamma functions one can read
the asymptotic behavior of the spectral function near $\omega=0$. Indeed for finite $k_2$
one finds ${\cal A}\sim \frac{k^2_2}{\omega}$ showing that it has a simple pole.

\begin{figure}
\begin{center}
\includegraphics[height=4.5cm, width=5cm]{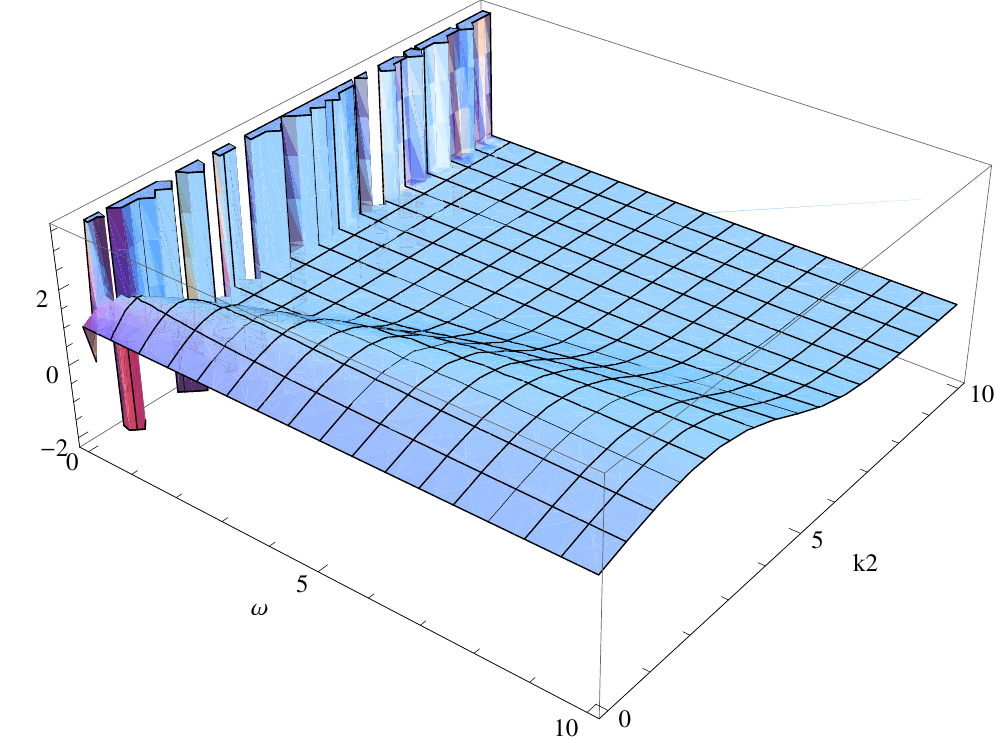}
\caption{The real part of the flat band eigenvalue, $\lambda_1$ which shows a delta function
behavior at $\omega=0$ indicating that the imaginary part of $\lambda_1$  has a pole at $\omega=0$.} \label{fig2}
\end{center}
\end{figure}

To further explore the physical content of the model it is useful to study the behavior of
the eigenvalues of the retarded Green's function as we approach $\omega=0$ for fixed
and finite $k_2$. Indeed using the asymptotic behavior of the Gamma functions,
for ${k_2^2}\gg \omega$, one finds
\be
\lambda_1=\frac{k_2-A}{k_2+A}\approx i\frac{k_2^2}{\omega},\;\;\;\;\;\;\;\;\;\;\;\;\;\;
\lambda_2=\frac{k_2+A}{k_2-A}\approx - i\frac{\omega}{k_2^2}.
\ee
This shows that for  finite values of $k_2$ one of the eigenvalues, $\lambda_1$,  has a pole
at $\omega=0$. More generally one can see that the eigenvalue $\lambda_1$ has a pole
at $\omega=0$ for all values of spatial momenta.
 This can be seen, for example, from the behavior of the real part of the
eigenvalue $\lambda_1$ where there is a delta function at $\omega=0$ as shown in figure
\ref{fig2}.  As a result one may conclude that there are localized non-propagating excitations
in the model showing that the theory exhibits a flat band.

Note that, as we have already mentioned in the introduction, the behavior of the retarded Green's function is
different from that considered in \cite{Faulkner:2010tq} where by making use of a semi-holographic method it 
was shown that the self energy of the fermions appeared in the Green's function is real. We note, however, due
to the fact we are using the Lorentz symmetry breaking boundary term the resultant Green's function has an 
imaginary part. Actually the situation is similar to the pure AdS case with Lorentz symmetry
breaking boundary term \eqref{boundary} studied in \cite{Tong}.
 Although in this case the bulk AdS  geometry respects the Lorentz symmetry,  the
boundary term breaks this symmetry leading to a non-relativistic boundary theory. On the other hand since the boundary term \eqref{boundary}, up to parity, is invariant under the Lifshitz
symmetry, the non-relativistic theory one gets from AdS bulk geometry has the same
symmetry as if we had started with Lifshitz geometry in the bulk.
Therefore one may conclude that the appearance of flat band is, indeed,
 the consequence of the non-relativistic feature of the dual theory.


\section{Finite temperature}

In this section we would like to redo the computations of the previous section for a
non-relativistic theory at finite temperature. Following the general idea of gauge/gravity duality
placing  the dual  theory at finite temperature corresponds to having a black hole in
the bulk gravity. Therefore in our case we should look for a black hole solution
in the asymptotic Lifshitz geometry. Actually black hole solutions in the
asymptotic Lifshitz geometry have been studied in
\cite{{Danielsson:2009gi},{Mann:2009yx},{Bertoldi:2009vn},{Balasubramanian:2009rx}}.
In particular the authors of \cite {Balasubramanian:2009rx} have analytically
 constructed  a black hole which asymptotes to a vacuum Lifshitz solution with $z=2$.
The solution  may be supported by different  actions with different field contents, though the metric has the same form as follows
\be\label{LifBH}
ds^2=-(1-\frac{r^2}{r_H^2})\frac{dt^2}{r^4}+\frac{dr^2}{r^2(1-\frac{r^2}{r_H^2})}
+\frac{d\vec{x}^2}{r^2}
\ee
where $r_H$ is the radius of horizon. The Hawking temperature is\cite{Balasubramanian:2009rx}
\be
T=\frac{1}{2\pi r_H^2}.
\ee
\subsection{Fermions on Lifshitz black hole}

Following our study in the previous section we will consider a neutral massless fermion
on the Lifshitz black hole given by equation \eqref{LifBH}. For this geometry
the non-zero components of vierbeins and spin connections are
\begin{eqnarray*}
(e_t)^a=\frac{r^2}{\sqrt{1-\frac{r^2}{r_H^2}}}\; \delta^{ta},\hspace{5mm}
(e_i)^a=r\delta^{ia},\hspace{5mm}
(e_r)^a=r\sqrt{1-\frac{r^2}{r_H^2}}\;\delta^{ra},
\end{eqnarray*}
and
\begin{eqnarray*}
(\omega_{tr})_a=-(\omega_{rt})_a=\left(\frac{2}{r^2}-\frac{1}{r_H^2}\right)\;
\delta_{ta},\hspace{5mm}
(\omega_{ir})_a=-(\omega_{ri})_a=-\sqrt{\frac{1}{r^2}-\frac{1}{r_H^2}}\;\delta_{ia}.
\end{eqnarray*}
Therefore the equation of motion for a massless fermion in this background,
setting
\be
\Psi=e^{i\omega t+ik\cdot x}\psi(r)=e^{i\omega t+ik\cdot x}
r^2\phi(r),
\ee
 reduces to
\be\label{feq}
\bigg[ \sqrt{1-\frac{r^2}{r_H^2}}\;\partial_r-\frac{r}{r_H^2\sqrt{1-\frac{r^2}{r_H^2}}}
(\frac{1}{2}-i\omega r_H^2 \Gamma^r\Gamma^t)+i\Gamma^rk\cdot \Gamma\bigg]\phi(r)=0.
\ee
To solve this equation it is useful to act by $D\ds$ on the first order equation to get
a second order differential equation. In fact, defining a new variable $x=\frac{r}{r_H}$, one finds
\be
(1-x^2)\frac{d^2\phi_\pm}{dx^2}-2x\frac{d\phi_\pm}{dx}
+\left[\nu(\nu+1)-\frac{\mu_\pm^2}{1-x^2}\right] \phi_\pm=0,
\ee
where $\phi_\pm=\frac{1}{2}(1\pm \Gamma^r\Gamma^t)\phi$, and
\be
\nu=-\frac{1}{2}+ir_H\sqrt{k^2+\omega^2r_H^2},\;\;\;\;\;\;\;
\mu_\pm=\pm\frac{1}{2}-i\omega r_H^2.
\ee
The resultant differential equation has a well known form whose solutions are
the associated  Legendre functions $P$ and $Q$. Therefore the most general solution
of the above equation is
\be\label{sol2}
\phi_\pm(r)=
c_1^\pm\; P(\nu,\mu_\pm, x)+c_2^\pm\; Q(\nu,\mu_\pm,x),
\ee
where $c^\pm_{1,2}$ are constant spinors.

Of course so far we have solved  the second order differential equation which its solution
is not necessarily a solution of the equation of motion which is a  first order differential
equation. In other words the constant spinors $c_{1,2}^\pm$ are not independent. In fact
in order to find a solution of the equation of motion one needs to plug the
solution \eqref{sol2} into  the equation of motion which in general leads to
certain relations between the constant spinors $c^{\pm}_{1,2}$. Indeed using the
recursion relations between the associated Legendre functions\cite{book,book2},
\bea
(1-x^2)\frac{dP(\nu,\mu_\pm,x)}{dx}&=&(\nu-\mu_\pm+1)(\nu+\mu_\pm)\sqrt{1-x^2}\;
P(\nu,\mu_\pm-1,x)\cr &&+\mu_\pm xP(\nu,\mu_\pm,x)\cr &&\cr
&=&-\sqrt{1-x^2}\;
P(\nu,\mu_\pm+1,x)-\mu_\pm xP(\nu,\mu_\pm,x),
\eea
one finds
\be\label{rel1}
c^-_{1,2}=ir_H\Gamma^r (k\cdot \Gamma) c^+_{1,2}.
\ee

In order to impose the ingoing boundary condition on the wave function at the horizon
we note that at near horizon the  oscillating part of the solution
has the following form
\be
\left(1-\frac{r}{r_H}\right)^{\pm i\omega r_H^2},
\ee
which in our notation the ingoing and outgoing waves correspond to
plus and minus signs, respectively. On the other hand using the asymptotic
behaviors of the associated Legendre functions near $x=1$ one observes that the function $Q$ has both the ingoing and the outgoing components, though the function
$P$ has only the ingoing part. Therefore in order to have a physical solution one needs
to set $c^\pm_2=0$. As a result the solution of the equation of motion of the
massless fermions on the Lifshitz black hole satisfying the ingoing boundary condition is
\be
\Psi^\pm=c^\pm r^2 e^{i\omega t+ik\cdot x}  P\left(\nu,\mu_\pm,\frac{r}{r_H}\right),
\ee
with $c^\pm$ being  constant two-component spinors
satisfying
\be\label{tt}
c^-=ir_H\Gamma^r (k\cdot \Gamma) c^+.
\ee

\subsection{Retarded Green's function}

In this subsection, using the solution we just found, we will compute
the retarded Green's function of a fermionic operator in the dual non-relativistic theory at finite temperature.
As we mentioned in the previous section in order to compute  the
corresponding retarded Green's function one needs to properly identify the source and
response of the dual operator which in turns depends on the boundary condition.  In this
section we will follow our notation in the previous
section and will consider the same boundary action
as that  given by the equation \eqref{boundary}. In this notation  the  source and the response
of the dual operator  are  given by
$(\chi_1,\zeta_2)$ and $(\chi_2,\zeta_1)$, respectively.

In order to read the proper  source and  response one  needs  to find
the asymptotic behavior
of the solution as we approach the boundary. In fact by making use of the asymptotic
behaviors of the associated Legendre functions (see for example \cite{book})
one gets
\be
\Psi_\pm\sim \frac{2^{\mu_\pm}\sqrt{\pi}}{\Gamma(\frac{1-\nu-\mu_\pm}{2})
\Gamma(1+\frac{\nu-\mu_\pm}{2})}\; r^2c^\pm e^{i\omega t+ik\cdot x}
\equiv  A_\pm r^2 c^\pm e^{i\omega t+ik\cdot x} .
\ee
 In other words one may write
\be
\Psi\sim r^2\begin{pmatrix}
A_-c_{\uparrow}^-\\
A_+c_{\downarrow}^+\\
A_+c_{\uparrow}^+\\
A_-c_{\downarrow}^-
\end{pmatrix}\;e^{i\omega t+ik\cdot x}.
\ee
Note also that in our notation, equation \eqref{tt} reads
\be
\begin{pmatrix}
c_{\uparrow}^-\\
0\\
0\\
c_{\downarrow}^-
\end{pmatrix}=-r_H
\begin{pmatrix}
k_1c_{\downarrow}^++k_2c_{\uparrow}^+\\
0\\
0\\
k_2c_{\downarrow}^+-k_1c_{\uparrow}^+
\end{pmatrix}
\ee

Altogether with this information the retarded Green's function of the dual fermionic operator
in the finite temperature non-relativistic theory is
\be
G_R(k)=-
\begin{pmatrix}
\frac{A_+^2+r_H^2k^2A_-^2+2r_Hk_2A_-A_+}{r_H^2k^2A_-^2-A_+^2}
&\frac{2r_Hk_1A_-A_+}{r_H^2k^2A_-^2-A_+^2}\\ &&\\
\frac{2r_Hk_1A_-A_+}{r_H^2k^2A_-^2-A_+^2} &
\frac{ A_+^2+r_H^2k^2A_-^2-2r_Hk_2A_-A_+}{r_H^2k^2A_-^2-A_+^2}
\end{pmatrix}.
\ee
It follows from this expression that $G_{11}(\omega,k_2)=G_{22}(\omega,-k_2)$ and
$\det(G_R)=-1$. The spectral function is also given by
\be
{\cal A}(k)=-\frac{1}{\pi}{\rm Im}({\rm Tr} G_R)=\frac{2}{\pi}{\rm Im}\bigg(
\frac{r_H^2k^2A_-^2+A_+^2}{r_H^2k^2A_-^2-A_+^2}\bigg).
\ee
To explore different features of  the retarded Green's function it is useful to use the
rotational symmetry to
set $k_1=0$. In this case  the retarded Green's function reads
\be
G_R(k_2)=-
\begin{pmatrix}
\frac{r_Hk_2A_-+A_+}{r_Hk_2A_--A_+}
&0\\ &&\\
0 &
\frac{r_Hk_2A_--A_+ }{A_++r_Hk_2A_-}
\end{pmatrix}.
\ee
Moreover for the spectral function one also finds
\be
{\cal A}(k_2)=\frac{2}{\pi}{\rm Im}\left(\frac{
\frac{r_H^2k_2^2}{4}\Gamma^2(\frac{1}{2}+X^+)\Gamma^2(\frac{1}{2}+X^-)
+\Gamma^2(1+X^+)\Gamma^2(1+X^-)}
{\frac{r_H^2k_2^2}{4}\Gamma^2(\frac{1}{2}+X^+)\Gamma^2(\frac{1}{2}+X^-)
-\Gamma^2(1+X^+)\Gamma^2(1+X^-)}\right),
\ee
with $X^\pm=\frac{i}{2}(\omega r_H^2\pm r_H\sqrt{k_2^2+\omega^2r_H^2})$.

It is instructive to study the behavior of the  spectral function in the small
temperature limit.  Physically small temperature means that we should look for the
energies much higher than the temperature, {\em i.e} $\frac{T}{\omega}\ll 1$.
Practically one may expand the above expression for $\omega r_H^2\gg 1$. Indeed
by making use of the asymptotic behaviors of the Gamma function,
up to order of ${\cal O}(\frac{T^2}{\omega^2})$, one arrives at
\bea
{\cal A}(k_2)&=&\frac{2}{\pi}{\rm Im}\bigg[
\frac{\Gamma^2(\eta+\frac{1}{2})+\Gamma(\eta)\Gamma(\eta+1)}
{\Gamma^2(\eta+\frac{1}{2})-\Gamma(\eta)\Gamma(\eta+1)}\cr &&\cr
&\times&
\left(1+\frac{i\pi T}{4\omega}\;\frac{(4\eta-1)\Gamma^2(\eta+\frac{1}{2})\Gamma(\eta)\Gamma(1+\eta))}
{\Gamma^4(\eta+\frac{1}{2})-\Gamma^2(\eta)\Gamma^2(\eta+1)}
\right)\bigg],
\eea
with $\eta=\frac{k_2^2}{4i\omega}$. As we see at leading order it is
exactly the same expression we have found  for the zero temperature case (see
the equation \eqref{A0}).

\begin{figure}
\begin{center}
\includegraphics[height=5cm, width=5cm]{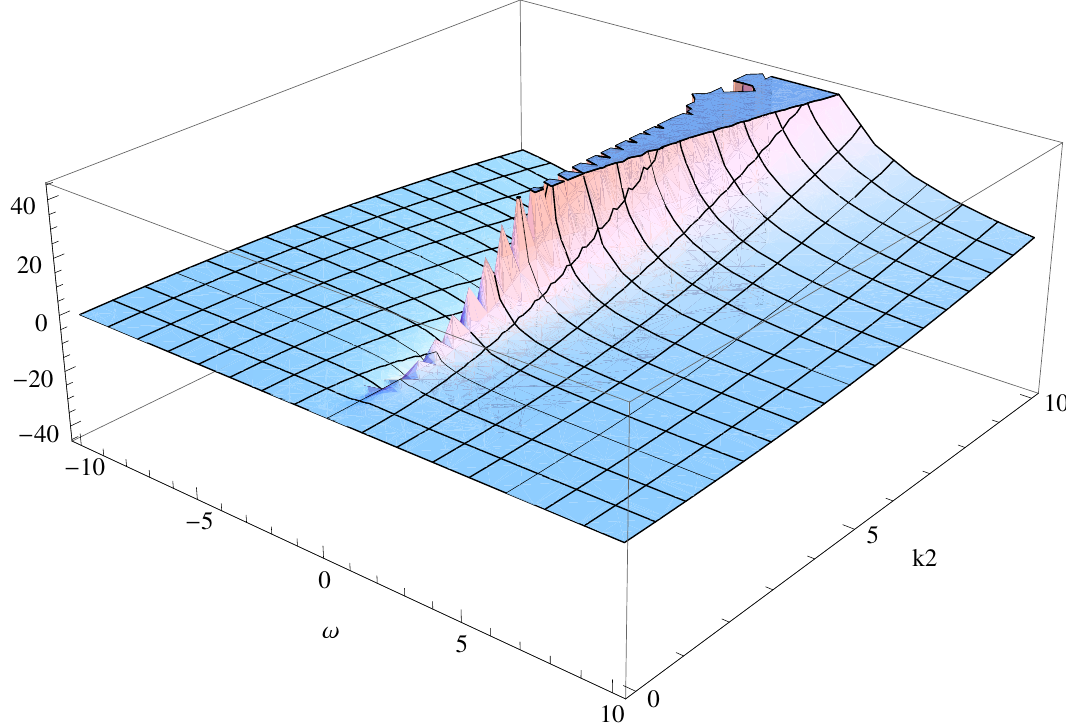}
\hspace{2cm}
\includegraphics[height=4.5cm, width=5cm]{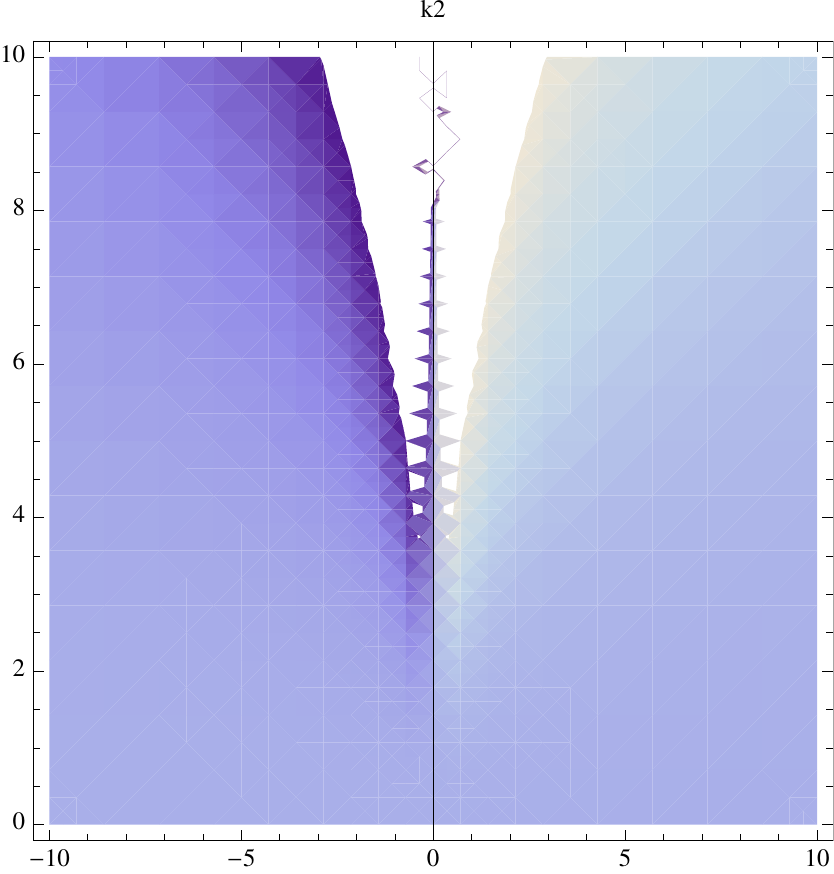}
\caption{Three dimensional and density plots of the spectral function
as a function of $\omega$ and $k_2$ at $r_H=\frac{2}{\sqrt{2\pi}}$. It is positive for $
{\rm sign}(\omega)>0$.
Note there is no pole
at $\omega=0$ for finite $k_2$.} \label{fig3}
\end{center}
\end{figure}

\begin{figure}
\begin{center}
\includegraphics[height=5cm, width=5cm]{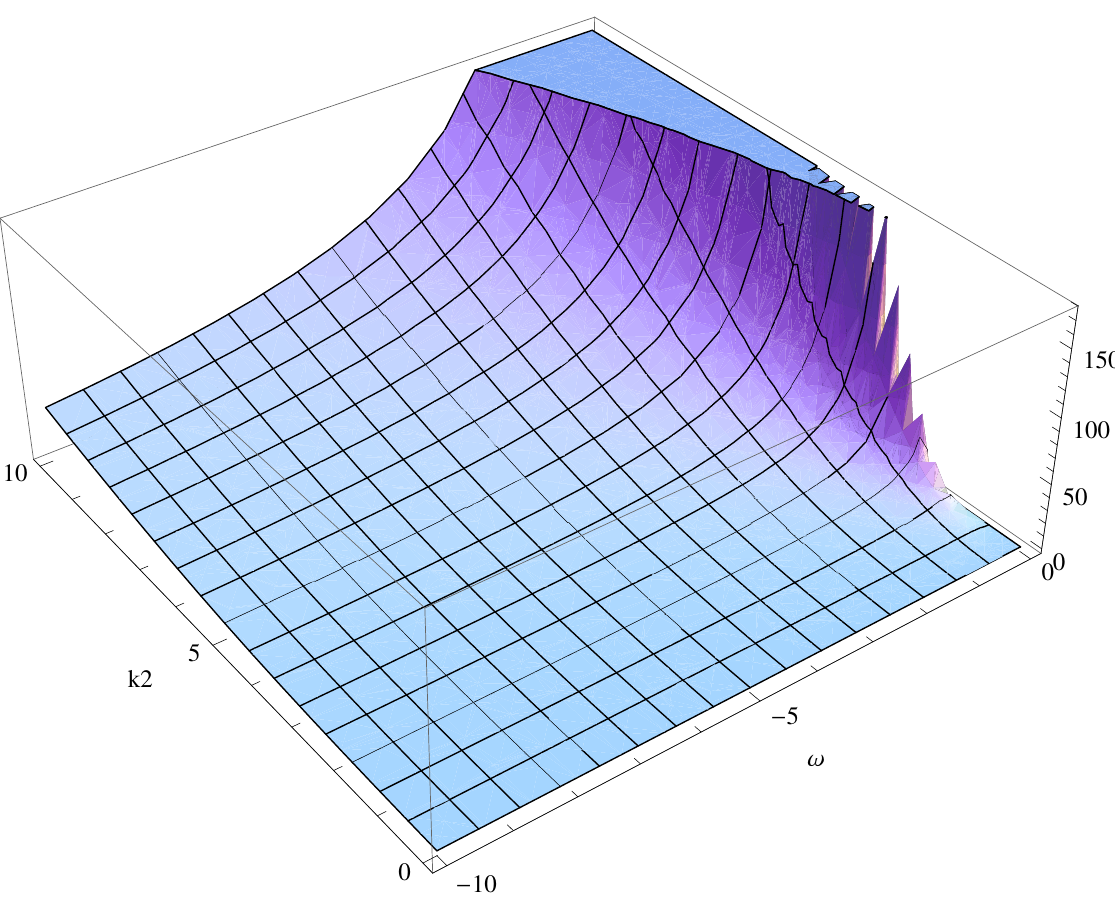}
\hspace{2cm}
\includegraphics[height=4.5cm, width=5cm]{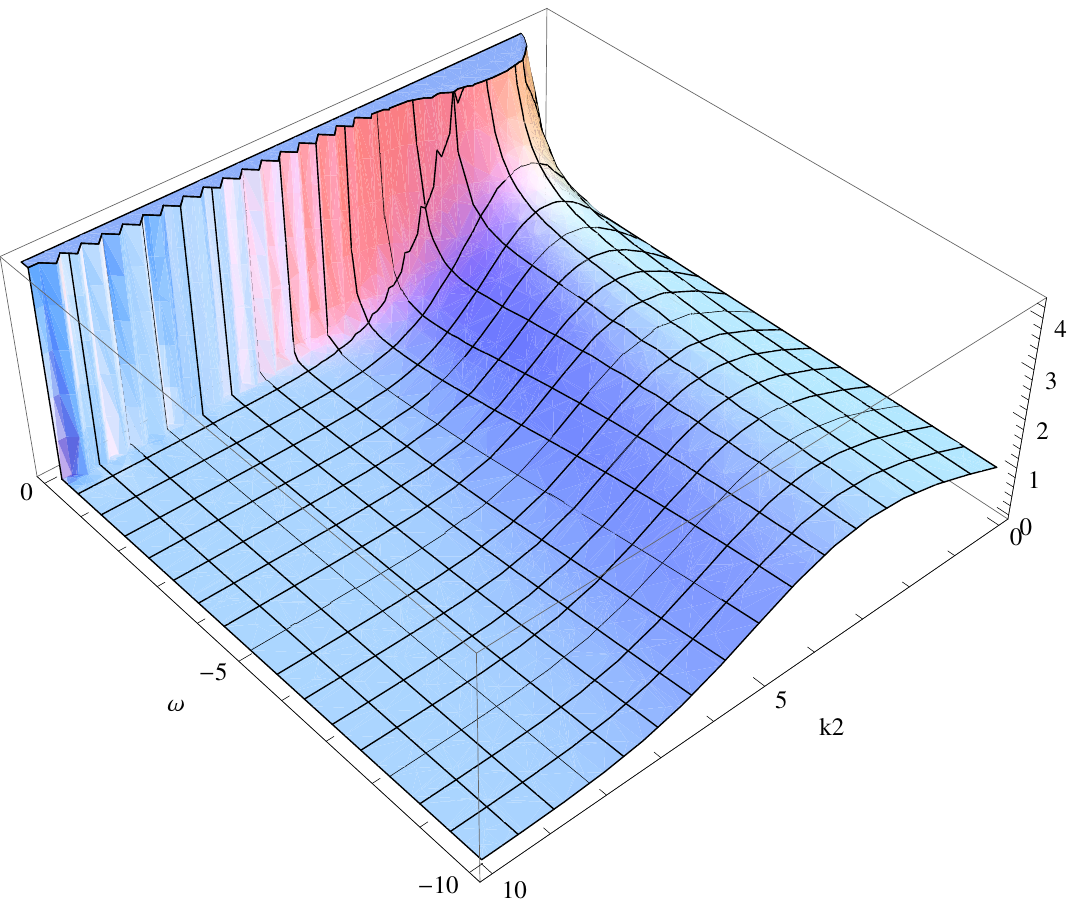}
\caption{The imaginary (left) and real (right) parts of the flat band eigenvalue.
 As we see
that there is a pole at $\omega=0$ in the imaginary part of the eigenvalue which is also
evident from the delta function behavior of its  real part.} \label{fig4}
\end{center}
\end{figure}

The spectral function as a function of $\omega$ and $k_2$ is depicted in figure \ref{fig3}. The plot is drawn
for $r_H=\frac{2}{\sqrt{2\pi}}$ where the temperature is $T=1/4$.
To further explore the physical content of the model it is also illustrative to examine the
behavior of the eigenvalues of the retarded
Green's function as functions of $\omega$ and $k_2$. In fact the real and imaginary parts
of the first  eigenvalue  of the retarded Green's function  have been plotted in figure
\ref{fig4}. As it is shown  the imaginary part
of the first eigenvalue has a pole at  $\omega=0$. This can also be seen from the
delta function behavior of its real part.  On the other hand it can be seen that the second eigenvalue has no pole at $\omega=0$.

In comparison with the zero temperature case we see that the spectral function has
qualitatively the same shape,  though there is a small deviation in the low momentum modes. Nevertheless for high momenta it remains unchanged.
Therefore the system has a flat band for high momenta.

It is worth to note that at low momenta and for low energies there are several non-trivial
peaks. In fact the presence of these peaks at low energies suggest that heating up
the system has excited zero energy fermionic modes at low momenta.

\section{Non-zero chemical potential}

It is important to mention that when one studies  fermionic features of a system
in  condensed matter physics,  usually one looks for a possibility of having a Fermi surface.

Actually in order to rigorously address this question one needs to
consider a charged fermion propagating on a charged Lifshitz black hole\footnote{
Few days after submitting our paper, another paper\cite{Fang:2012pw}
appeared on arXiv where
charged fermions on the Lifshitz geometry were studied. Of course their background is
different form what we are considering in this paper for charged fermions.}
where we could have a non-zero chemical potential.
In fact as we have already mentioned in section two  the Lifshitz geometry
is not a solution of
pure Einstein gravity. In order to find the Lifshitz solution one may couple gravity to a
 massive background gauge field. In this case the background supports a non-zero
gauge field.

We note, however, that this gauge field diverges as we approach the boundary and thus
cannot play the role of chemical potential. In fact  in order to have a chemical  potential,
another gauge field is needed \cite{Tarrio:2011de}. Indeed the second gauge field has the proper
near boundary behavior to define  chemical potential. More precisely one may start with the following action\cite{Tarrio:2011de}
\be
S=\frac{1}{16\pi G_4}\int d^4x\sqrt{-g}\bigg[R-2\Lambda-\frac{1}{2}(\partial\phi)^2-
\frac{1}{4}e^{\lambda_1\phi} ({F^{(1)}})^2-\frac{1}{4} e^{\lambda_2\phi}({ F^{(2)}})^2\bigg].
\ee
This model admits a charged Lifshitz black hole solution with critical exponent $z$ for the
particular values of $\Lambda, \lambda_1$ and $\lambda_2$ as follows
\be
\Lambda=-\frac{(z+1)(z+2)}{2L^2},\;\;\;\;\;\lambda_1=-\frac{2}{\sqrt{z-1}},
\;\;\;\;\;\lambda_2=\sqrt{z-1}.
\ee
The corresponding black hole solution is\cite{Tarrio:2011de}\footnote{
Note that in comparison with the solution in \cite{Tarrio:2011de} we have shifted
the gauge field by constants to make sure that $g^{\mu\nu}A^{(i)}_\mu A^{(i)}_\nu$ remains finite. Moreover by a proper rescaling we have set $L=1$ and also with a proper choice of the
parameters the radius of horizon has also been set to one.}
\bea
&&ds^2=-r^{2z}fdt^2+\frac{dr^2}{r^2f}+r^2d\vec{x}^2,\;\;\;\;\;\;
\;\;\;\;\;
f=1-\frac{1+r_0^{2(z+1)}}{r^{z+2}}+\frac{r_0^{2(z+1)}}{r^{2(z+1)}},
\cr
&&e^{\sqrt{z-1}\phi}=
\frac{\kappa^2 }{4zr_0^{2(z+1)}}r^{2(z-1)},\cr
&&
A^{(1)}_t=-\mu^{(1)}\bigg(1-r^{2+z}\bigg),\;\;\;\;
A^{(2)}_t=\mu^{(2)}\bigg(1-\frac{1}{r^{z}}\bigg),
\eea
where $r_0$ and $\kappa$ are the only remaining free parameters which determine mass and
charges of the solution, and
\be
\mu^{(1)}=\sqrt{\frac{2(z-1)}{z+2}}\left(\frac{\kappa^2} {4zr_0^{2(z+1)}}\right)^{\frac{1}{z-1}},\;\;\;\;\;\;\;\;\;\;\;\;
\mu^{(2)}=\frac{4r_0^{2(z+1)}}{\kappa }.
\ee
 It is, indeed, an asymptotically Lifshitz charged black hole whose
Hawking temperature is
\be
T=\frac{z+2}{4\pi}\left(1-\frac{z}{z+2}r_0^{2(z+1)}\right).
\ee

At low energy, using the general idea of AdS/CFT correspondence,
the physics is governed by near horizon modes. In this case, for example,  one should study fermions
on the near horizon background.  At zero temperature where
$r_0^{2(z+1)}=\frac{z+2}{z}$ setting
\be \label{scaling}
r-1=\frac{\epsilon }{(2+3z+z^2)  \xi},\;\;\;\;\;\;\;\;\;t=\frac{1}{\epsilon} \tau,
\ee
the near horizon background can be obtained in the limit of $\epsilon\rightarrow 0$
where one finds
\bea\label{AdS2}
ds^2=\frac{-d\tau^2+d\xi^2}{(2+3z+z^2)\xi^2}+ dx_2^2,\;\;\;\;
e^{\phi}=\frac{\kappa^2}{4(z+2) },\nn\\
A^{(1)}_\tau=\frac{\mu^{(1)} }{(z+1)\xi},\;\;\;\;\;A^{(2)}_\tau=\frac{z \mu^{(2)} }{(z^2+3z+2)\xi}.
\eea
As we observe the metric is in the form of $AdS_2\times R^2$. Therefore the low energy physics is described by an emergent IR CFT. Actually the situation is very similar to the relativistic
case \cite{Faulkner:2009wj}. As a result we would expect that the model exhibits
a Farmi surface whose physics is governed by an IR fixed point.

\subsection{Charged Fermions}

Let us first consider a four dimensional charged Dirac fermion on the Lifshitz background whose
action is
\be S_{\rm bulk} =\int d^4x \sqrt{-g}\,i\bar{\Psi} \,\left[\frac{1}{2}\left(\Gamma^a\!\stackrel{\rightarrow}D_a-
\stackrel{\leftarrow}D_a\!\Gamma^a\right)-m\right]\Psi\label{Action}.
\ee
Here
${D\ds}=(e_\mu)^a\Gamma^\mu[\partial_a+\frac{1}{4}(\omega_{\rho\sigma})_a\Gamma^{\rho\sigma}-iq A^{(2)}_a]
$, with
$\Gamma^{\mu\nu}=\frac{1}{2}[\Gamma^\mu,\Gamma^\nu]$.

As we discussed in the previous sections one needs to impose a proper boundary condition
to get a well defined  variational principle. With a suitable boundary condition the
equation of motion is
\be
\left((e_\mu)^a\Gamma^\mu\left[\partial_a+\frac{1}{4}(\omega_{\rho\sigma})_a\Gamma^{\rho\sigma}-iqA^{(2)}_a\right]-m\right)\Psi=0.
\ee
The above equation of motion by the choice of $\Psi=(-h)^{-1/4}e^{-i\omega t+ik.x}\psi(r)$ reduces to
\be\label{EOM}
\bigg[rf^{1/2}\Gamma^r\partial_r-\frac{i}{r^zf^{1/2}}\left(\omega+q \mu^{(2)}(1-\frac{1}{r^{z}})\right)
\Gamma^t+\frac{i }{r} \Gamma\cdot k-m\bigg]\psi(r)=0.
\ee

As we already mentioned the low energy is governed by an emergent IR fixed point. To
examine the low energy limit of the fermions one should consider the limit of
$\omega \ll \mu^{(2)}$. At zero temperature, using the
scaling \eqref{scaling} and in the limit of $\epsilon\rightarrow 0$ keeping $\omega/\epsilon$
fixed the above equation reads
\be
\left[-\Gamma^\xi\xi\partial_\xi-i\xi\Gamma^t(\tilde{\omega}+\frac{qe}{\xi})
+i\Gamma\cdot{k}-\frac{ m}{\sqrt{2+3z+z^2}}\right]\psi(\xi)=0,
\ee
where
\be
e=\frac{(1+z)\mu^{(2)}}{2+3z+z^2},\;\;\;\;\;\;\;\;
\;\;\;\;\;\;\;\;\tilde{\omega}=\frac{ \omega}{\epsilon}={\rm finite}.
\ee

We recognize the above equation as a charged fermion probing the $AdS_2\times R^2$
background \eqref{AdS2}, as expected. As a result one may go through the construction
of \cite{Faulkner:2009wj} to express the retarded Green's function of the fermion
on the Lifshitz charged black hole in terms of the
retarded Green's function of $AdS_2$ model in small $\omega$ limit.

Here instead of doing so, one utilizes the numerical method to solve the equation of
motion numerically. To proceed, it is useful to consider  the following representation for four dimensional gamma matrices
\begin{eqnarray}\label{basis2}
\Gamma^r=
\begin{pmatrix}
-\sigma^3&0\\0&-\sigma^3
\end{pmatrix},
\Gamma^t=
\begin{pmatrix}
i\sigma^1&0\\0&i\sigma^1
\end{pmatrix},
\Gamma^1=
\begin{pmatrix}
-\sigma^2&0\\0&\sigma^2
\end{pmatrix},
\Gamma^2=
\begin{pmatrix}
0&-i\sigma^2\\i\sigma^2&0
\end{pmatrix}.
\end{eqnarray}

Due to rotational symmetry in the spatial directions we may set $k_2=0$. Then using
the notation
\begin{eqnarray}
\psi=
\begin{pmatrix}
\Phi_1\\ \Phi_2
\end{pmatrix},
\end{eqnarray}
the equation of motion \eqref{EOM} reduces to the following decoupled equations
\be\label{Eq}
\bigg[rf^{1/2}\partial_r-\frac{1}{r^{z}f^{1/2}}\left(\omega+q \mu^{(2)}(1-\frac{1}{r^{z}})\right)
i\sigma^2+m\sigma^3-(-1)^\alpha\frac{k_1}{r} \sigma^1\bigg]\Phi_\alpha=0,
\ee
for $\alpha=1,2$.  It is easy to see that
\be
\Phi_\alpha\sim a_\alpha r^{m}\begin{pmatrix} 0\\ 1\end{pmatrix}+
 b_\alpha r^{-m}\begin{pmatrix} 1\\ 0\end{pmatrix},\;\;\;\;\;\;\;\;\;{\rm for}\;r\rightarrow
\infty.
\ee

To find the retarded Green's function, following \cite{Faulkner:2009wj}, it is useful
to defined $\zeta_1=\psi_1/\psi_2$ and $\zeta_2=\psi_3/\psi_4$ where
$\psi_i$'s are defined via $\Phi_1=(\psi_1, \psi_2), \Phi_2=(\psi_3,\psi_4)$.
These parameters satisfy the
following equations
\bea
&&{rf^{1/2}}\partial_r\zeta_1+2m\zeta_1-\left(\frac{\Omega}{r^zf^{1/2}}+
\frac{k_1}{r} \right) \zeta_1^2=\frac{\Omega}{r^zf^{1/2}}-\frac{k_1}{r} ,\cr &&\cr
&&{rf^{1/2}}\partial_r\zeta_2+2m\zeta_2-\left(\frac{\Omega}{r^zf^{1/2}}-
\frac{k_1}{r} \right) \zeta_2^2=\frac{\Omega}{r^zf^{1/2}}+\frac{k_1}{r} ,
\eea
where
\be
\Omega=\omega+q \mu^{(2)}(1-\frac{1}{r^{z}}).
\ee
Using these equations the  retarded Green's function is essentially given in terms
of functions $G_1(k,\omega)$ and $G_2(k,\omega)$ where\footnote{Here we set $k_1=k$.}
\be
G_\alpha(k,\omega)=\lim_{r\rightarrow \infty} r^{2m} \zeta_\alpha,\;\;\;\;\;\;\;\;\;\;
{\rm for}\;\alpha=1,2.
\ee
with the
ingoing boundary condition at the horizon which in our notation it is  \cite{Faulkner:2009wj}
\be
\zeta_\alpha|_{\rm horizon}=i.
\ee

The precise expression of the retarded Green's function in terms of $G_\alpha$
depends on the boundary condition  one imposes to
get a well defined variational principle. For example, if we impose the standard
boundary condition, in our notation, the corresponding retarded Green's function is
\be
G(k,\omega)= -
\begin{pmatrix}
G_1(k,\omega)&0\\0&G_2(k,\omega)
\end{pmatrix}.
\ee
Therefore the spectral function reads
\be
{\cal A}(k,\omega)=\frac{1}{\pi}{\rm Im} \bigg(G_1(k,\omega)+G_2(k,\omega)\bigg).
\ee

On the other hand for the boundary condition obtained by adding the boundary
term \eqref{boundary} the corresponding retarded Green's function  as a function of
$G_\alpha$ is given by (see also \cite{Li:2011nz})
\be
G(k,\omega)= -
\begin{pmatrix}
\frac{2G_1(k,\omega)G_2(k,\omega)}{G_1(k,\omega)+G_2(k,\omega)}&
\frac{G_1(k,\omega)-G_2(k,\omega)}{G_1(k,\omega)+G_2(k,\omega)}\\ &\\
\frac{G_1(k,\omega)-G_2(k,\omega)}{G_1(k,\omega)+G_2(k,\omega)}&
\frac{-2}{G_1(k,\omega)+G_2(k,\omega)}
\end{pmatrix}.
\ee
Thus the corresponding spectral function reads
\be
{\cal A}(k,\omega)=2{\rm Im}\bigg(\frac{G_1(k,\omega)G_2(k,\omega)-1}
{G_1(k,\omega)+G_2(k,\omega)}\bigg).
\ee

Note that in the notation we are using in this section (see
\eqref{basis2}) the boundary term \eqref{boundary} reads
\bea\label{boundary}
S_{bdy}=\frac{i}{2}\int d^3x\sqrt{-h}\bar{\Psi}\Gamma^1\Gamma^2\Psi
=\frac{-i}{2}\int d^3x\sqrt{-h}(\psi^\dagger_1\psi_4+\psi^\dagger_2\psi_3+
\psi^\dagger_3\psi_2+\psi^\dagger_4\psi_1).
\eea
Therefore adding this boundary term to the action results to impose the boundary condition on a
combination of the different components of the fermions as follows
\be
\delta S=i\int d^3x\sqrt{-h}
(\delta\chi_2^\dagger \eta_2-\delta\eta_1^\dagger
\chi_1-\chi_1^\dagger\delta\eta_1+\eta^\dagger_2\delta\chi_2),
\ee
where
\bea
(\chi_1,\chi_2)&=&\frac{1}{\sqrt{2}}(\psi_2+\psi_4,\psi_2-\psi_4),\cr
(\eta_1,\eta_2)&=&\frac{1}{\sqrt{2}}(\psi_1+\psi_3,\psi_1-\psi_3),
\eea

\subsection{Numerical results}

\begin{figure}
\begin{center}
\includegraphics[height=5.5cm, width=6cm]{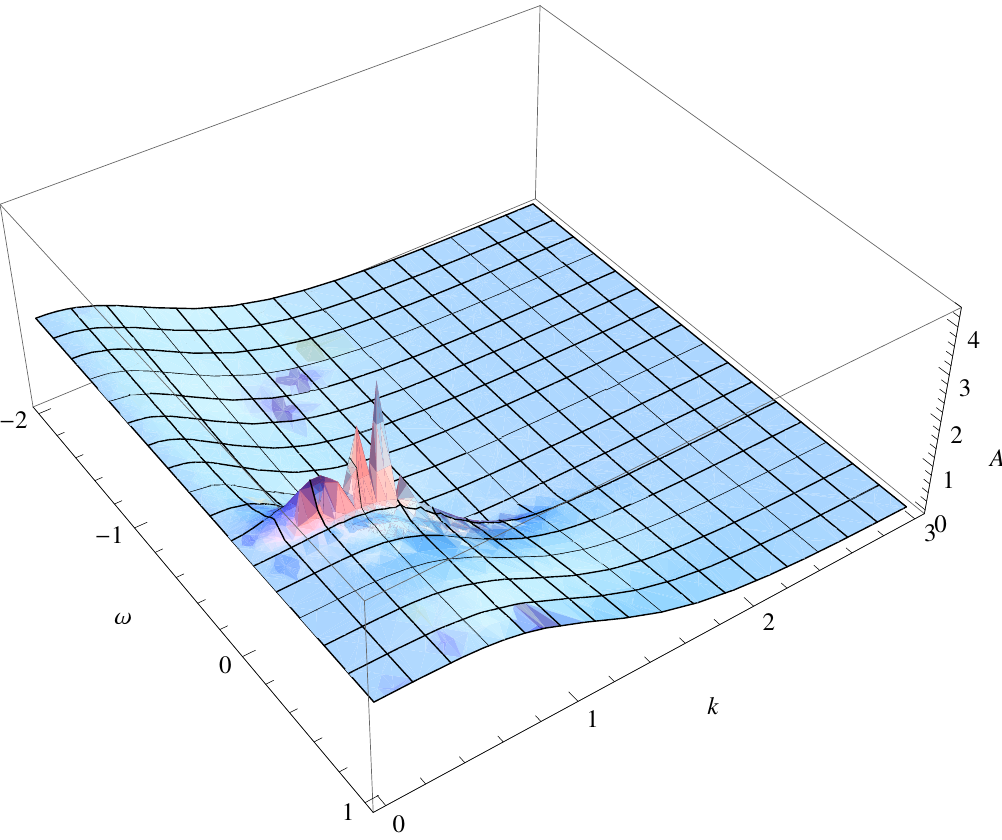}
\hspace{1cm}
\includegraphics[height=5.5cm, width=6cm]{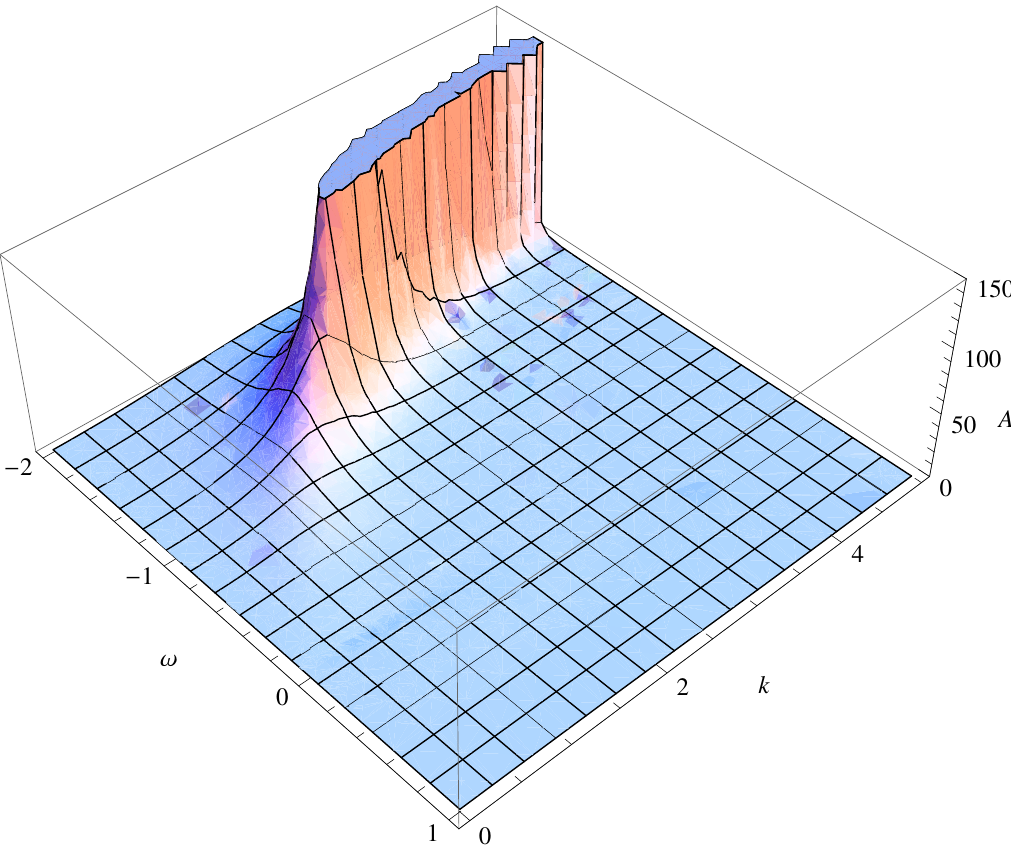}
\caption{The behavior of spectral functions with the standard (left) and non-standard (right) boundary conditions
for $z=2$  and $T=0$ as  functions of $\omega$ and $k$. For the standard boundary condition the system exhibits a
Fermi surface at $k_f=0.8902$, while for the non-standard case it has a flat band.}\label{fig5}
\end{center}
\end{figure}

Having found  expressions for the retarded Green's function and the spectral function
for the cases of standard and non-standard boundary conditions, it is an easy task to
find their behaviors as functions of $k$ and $\omega$. Here to
explore the physical content of the model we have  plotted the spectral function
of the model for both standard and non-standard boundary conditions.
At zero temperature where $r_0^{2(z+1)}=(z+2)/z$ we set
$m=0$, $q\mu^{(2)}=\sqrt{3}$. For $z=2$ the spectral function is shown in figure
\ref{fig5}. While in the standard case the model has a Fermi surface at $k_f=0.8902$,
in the non-standard case it exhibits a flat band. Note that since the
retarded Green's function is an even function of $k$ (see equations \eqref{Eq}) we only  considered $k>0$.

It is worth to note that the model contains four free parameters which are
mass $m$, critical exponent $z$, temperature $T$ and chemical potential $\mu^{(2)}$ which
always appears in the  combination $q\mu^{(2)}$ where $q$ is the charge of the fermion.
Therefore it is natural to explore the physical content of the model when we vary these
parameters. Actually changing the parameters we find the following behaviors.

\begin{figure}
\begin{center}
\includegraphics[height=5.5cm, width=6cm]{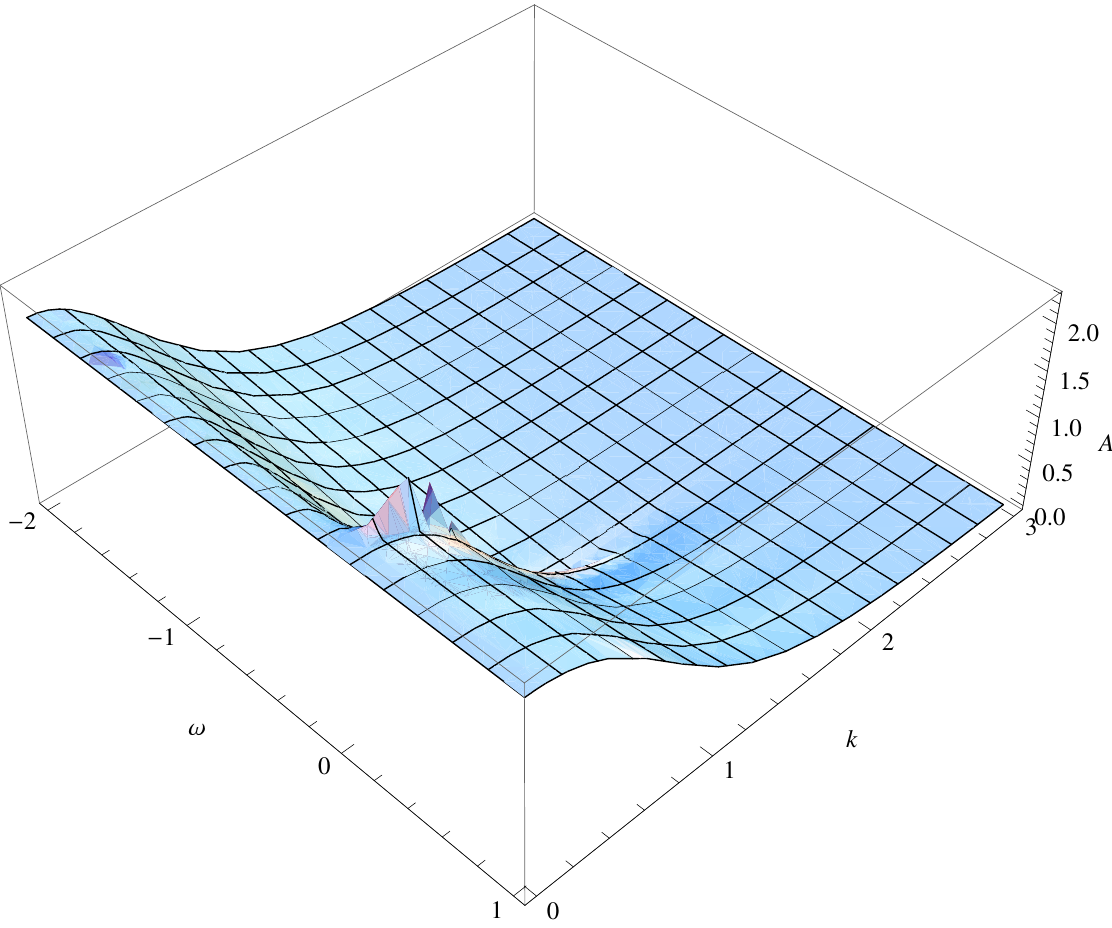}
\hspace{1cm}
\includegraphics[height=5.5cm, width=6cm]{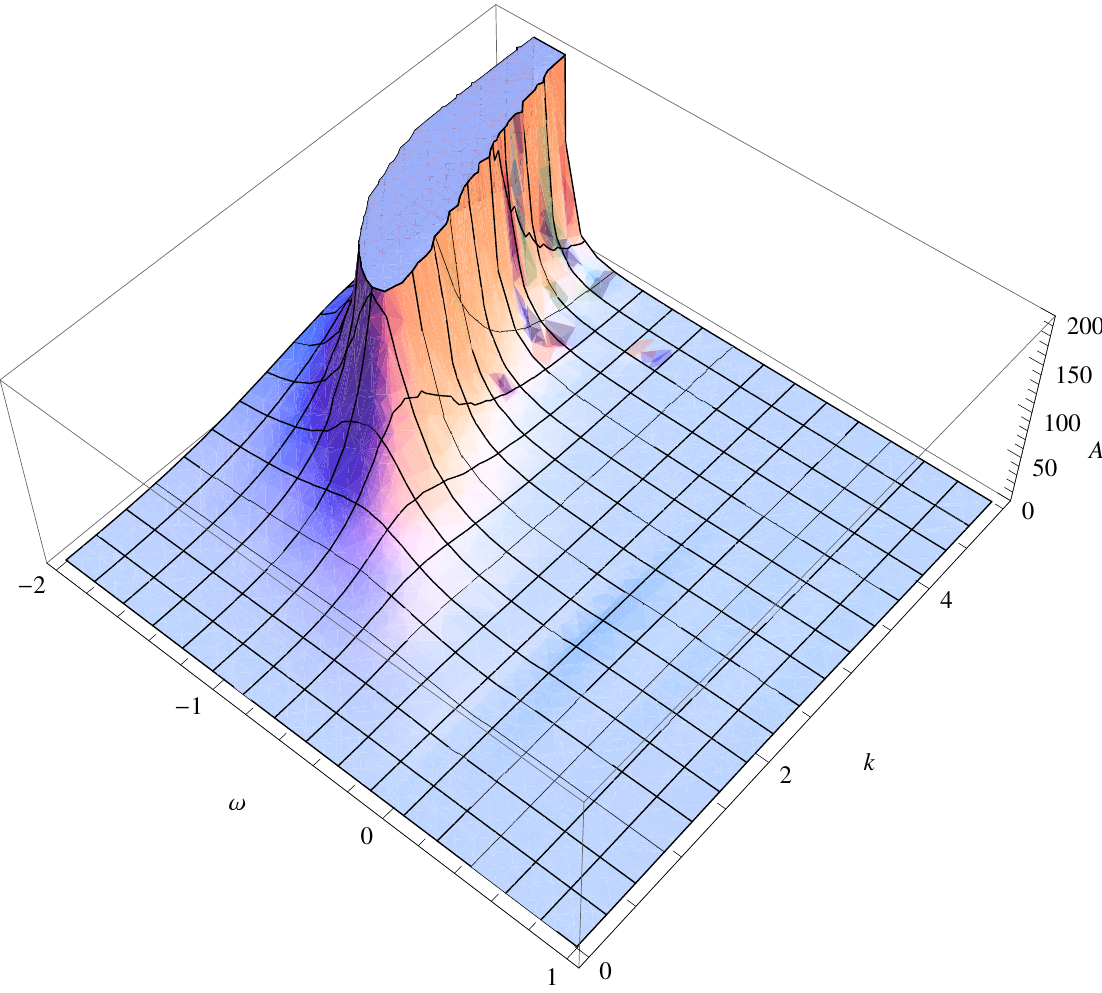}
\caption{The behavior of spectral functions with the standard (left) and non-standard (right) boundary conditions
for $z=3$  and $T=0$ as  functions
of $\omega$ and $k$. For the standard boundary condition the system exhibits a
Fermi surface
, while for the non-standard case it has a flat band.}\label{fig6}
\end{center}
\end{figure}

For fixed $m,T$ and $q\mu^{(2)}$ as we increase the critical exponent $z$
for the standard boundary condition the sharp peak representing the Fermi surface  becomes smaller and occurs at smaller $k_f$ and eventually for large enough $z$ it
 distorts the Fermi surface completely. In other words the model does not have
a Fermi surface. On the other hand for the non-standard boundary condition
the model still exhibits a flat band  though there is a depletion in the low momentum modes
(see for example figure \ref{fig6} for $z=3$).

For fixed $z$ the dependence of the spectral function on the parameters $m,T$ and $q\mu^{(2)}$
for both standard and non-standard cases is the same as that for $z=1$ where we have
AdS black hole solutions (see for example \cite{Faulkner:2009wj} and \cite{Tong}
for standard and non-standard boundary conditions, respectively).
\section{Discussions}

In this paper, following the general idea of AdS/CFT correspondence,  we have 
studied  retarded Green's function of a fermionic operator in a three dimensional 
non-relativistic field theory
by making use of a massless fermion on the asymptotically Lifshitz geometry. 
 In this paper we have mainly considered the asymptotically
Lifshitz geometry with critical exponent of $z=2$. We have considered 
both neutral and charged fermions. 

Taking into account that 
the gravity on asymptotically Lifshitz backgrounds may provide  holographic descriptions for 
strange metals \cite{Hartnoll:2009ns} our studies might be useful to explore certain 
features of strange metals.

For neutral fermions at zero temperature where the bulk fermions propagate on the Lifshitz
background the resultant retarded Green's function of the dual fermionic
operator  exhibits interesting behaviors.
In particular we observe that the spectral function has a pole at $\omega= 0$ for all values 
of spatial momenta. The appearance of the pole may also be seen from the behavior of the 
real and imaginary parts of the eigenvalues of the corresponding retarded Green's function.

Having seen a pole at $\omega=0$ with all values of spatial momenta shows that 
there are localized non-propagating excitations which in turns indicates  an infinite
flat band. Actually the  situation is similar to that 
in pure AdS geometry with the Lorentz breaking boundary condition\cite{Tong}.

We have also considered three dimensional non-relativistic theories at finite temperature.
To study the three dimensional model at finite temperature we have utilized the asymptotically 
Lifshitz  black hole obtained in  \cite{Balasubramanian:2009rx}. We have shown that by heating
up the dual theory, although the non-zero temperature can excite low momenta zero energy
modes, at high momenta there is still an infinite flat band!

An interesting feature we have seen in our model is that the spectral function 
is positive for ${\rm sign}(\omega)>0$. In fact unitarity requires that the spectral function
to be always positive. Since in our case the retarded Green's function changes its
sign and indeed  is negative for  ${\rm sign}(\omega)<0$, it is tempting to propose
that retarded Green's function we have found for the non-relativistic model contains information
for both particles and anti-particles! 

We have also considered charged fermions probing a charged Lifshitz black hole. While 
for the standard boundary condition the model exhibits a Fermi surface for 
non-standard boundary condition the model still has a falt band.
We have also observed that as one increases the critical exponent $z$
for the standard boundary condition the sharp peak representing the Fermi surface becomes smaller and occures at smaller $k_f$ and eventually for large enough $z$ it 
 destroies the Fermi surface completely. In other words the model does not have
a Fermi surface. On the other hand for the non-standard boundary condition 
the model still exhibits a flat band  though there is a depletion in the low momentum modes.

It is worth to note that in order to make the variational principle well defined one could also use another boundary action as follows
\be
S_{bdy}=\frac{1}{2}\int d^3x\sqrt{-h}\;(\bar{\Psi}_+\Psi_--\bar{\Psi}_-\Psi_+)=\frac{i}{2}\int d^3x\sqrt{-h}(\Psi_2^\dagger\Psi_1+\Psi_3^\dagger\Psi_4-
\Psi_1^\dagger\Psi_2-\Psi_4^\dagger\Psi_3).
\ee
With this boundary term the variation of the whole action leads to the following boundary terms 
\be
i\int\sqrt{-h}\left[-\delta \chi^{\dagger}_2\chi_1+\delta \zeta^{\dagger}_2\zeta_1+ \chi_1^{\dagger}\delta \chi_2-\zeta_1^{\dagger}\delta \zeta_2\right],
\ee
where in this case the newly defined fields are given by
\bea
(\chi_1,\chi_2)&=&\frac{1}{\sqrt{2}}(\Psi_1+\Psi_2,\Psi_1-\Psi_2)\cr
(\zeta_1,\zeta_2)&=&\frac{1}{\sqrt{2}}(\Psi_3+\Psi_4,\Psi_3-\Psi_4).
\eea
If we follow the steps we went through in the previous sections one  may 
compute the corresponding  retarded Green's function in this case. Doing so, 
one finds that the resultant
retarded Green's functions have the same form as those in the previous sections,
except the fact that the roles of  $k_1$ and $k_2$ have been changed. 
Now in this case we could use the rotational symmetry to
set $k_2=0$. Of course the physics remains unchanged after all.

{\bf Note added:} After submitting our paper we were informed by U. Gursoy
that fermionic correlation functions on Lifshitz background has recently been studied in \cite{Gursoy:2011gz}. We note, however, that the authors of this paper have used different 
UV boundary condition than ours.

\section*{Acknowledgments}

We would like to thank Davoud Allahbakhshi, Ali Davody, Reza Fareghbal, Umut Gursoy, Joao N. Laia and  David Tong  for
useful comments and  discussions. M. A. would like to thank CERN TH-division
for hospitality. This work is supported by Iran National Science Foundation (INSF).

\end{document}